\documentclass[aps,prl,reprint,superscriptaddress,showpacs,amsmath,amssymb]{revtex4-2}
\usepackage{xr}
\usepackage{graphicx}
\usepackage[utf8]{inputenc}
\usepackage{float}
\usepackage{dcolumn}
\usepackage{bm}
\usepackage{lineno}
\usepackage{rotating}
\usepackage{url}
\usepackage{braket}
\usepackage{mathrsfs}
\usepackage{overpic}

\def\ba{\begin{eqnarray}}
\def\ea{\end{eqnarray}}

\usepackage[colorlinks,citecolor=blue,linkcolor=blue,breaklinks=true]{hyperref}

\begin{document}

\title{Universal observable as a signal of chiral anomaly in lattice Weyl fermions}

\author{Shi Chen}
\affiliation{Graduate School of China Academy of Engineering Physics, Beijing 100193, China}

\author{Yu Chen}
\email{ychen@gscaep.ac.cn}
\affiliation{Graduate School of China Academy of Engineering Physics, Beijing 100193, China}

\begin{abstract}
The Adler-Bell-Jackiw chiral anomaly is shown to retain its Lorentz-invariant form, $\partial_\mu J^\mu_5 \propto \mathbf{E} \cdot \mathbf{B}$, in lattice Weyl systems beyond moderate magnetic fields, where neither Lorentz nor rotational symmetry is present. We show that the longitudinal and Hall magnetoconductivities factorize into a product of a universal part, governed by the chiral anomaly, and a non-universal part that depends on the density of states at the Fermi level. A rotationally invariant observable $\varkappa = \sigma (c_V/T)^2$ is introduced as a robust signature of the anomaly, where $\sigma$ denotes the Euclidean norm of the longitudinal and Hall conductivities and $c_V$ is the specific heat density. This quantity follows a universal $B^2$ dependence and scales as $|\cos\Theta|$, with $\Theta$ being the angle between $\mathbf{E}$ and $\mathbf{B}$. Through analytical derivation and full numerical simulation, we establish that $\varkappa$ remains universal independent of system parameters and of the orientation of the magnetic or electric field for fixed $\Theta$. The emergent SO(3) symmetry in $\varkappa$ persists despite the absence of isotropy in both the microscopic model and the low-energy effective theory.
\end{abstract}

\maketitle

The Adler-Bell-Jackiw (ABJ) chiral anomaly is a fundamental quantum phenomenon in which a classically conserved chiral current fails to remain conserved upon quantization \cite{Adler1969, BellJackiw1969}. For a massless Dirac fermion, the axial current $J_5^\mu$, associated with the symmetry $\psi \rightarrow e^{i\alpha\gamma_5}\psi$, satisfies
\ba
\partial_\mu J_5^\mu = \frac{e^2}{16\pi^2\hbar^2} \epsilon^{\mu\nu\rho\sigma} F_{\mu\nu}F_{\rho\sigma} = \frac{e^2}{2\pi^2\hbar^2} \mathbf{E} \cdot \mathbf{B},\nonumber
\ea
where the divergence is proportional to the topological density $\mathbf{E} \cdot \mathbf{B}$. This relation governs processes such as neutral pion decay and underlies a range of topological effects in condensed matter systems. Crucially, the form of the anomaly is exact under radiative corrections (up to renormalization of the fine-structure constant) and is insensitive to fermion masses—a result known as the Adler-Bardeen theorem \cite{AdlerBardeen1969}. This robustness lies at the heart of many topological aspects of quantum field theory and motivates a deeper investigation into its stability, particularly in systems where Lorentz symmetry is absent.

The recent discovery of emergent Weyl fermions in condensed matter systems \cite{wan11, Lv2015a, Lv2015b, Xu2015a, Xu2016, Yang2015} offers a new platform for exploring the chiral anomaly in the presence of Lorentz symmetry breaking. Early work by Nielsen and Ninomiya identified negative magnetoresistance as a key signature of the anomaly in lattice systems \cite{NielsenNinomiya1983}. Subsequent semiclassical analyses by Son, Spivak, Burkov, and others established a quadratic-in-B negative magnetoresistance \cite{SonSpivak2013, Burkov2014}. These results, however, assume long-range disorder, which enforces a separation of timescales between intravalley and intervalley scattering and thereby supports steady-state currents. In experiments, the magnetotransport behavior is often more complex, exhibiting strong angular anisotropy deviating from semiclassical predictions or showing unconventional magnetic-field dependence \cite{Zhang2016, Xiong2015}. As noted by Das Sarma et al., short-range disorder can substantially modify the magnetoresistance, potentially leading to a constant or other non-universal B dependence \cite{DSarma2015}. These observations suggest that longitudinal magnetoconductivity alone is not a definitive probe of the chiral anomaly.

In this work, we address two key questions: whether the anomaly form itself is modified by the absence of Lorentz and rotational symmetry in lattice systems, and whether a more robust experimental signature can be identified. We show that the anomaly equation $\partial_\mu J_5^\mu \propto \mathbf{E} \cdot \mathbf{B}$ remains valid even in the presence of strong lattice effects, provided the system is in the quantum limit where only the chiral Landau level is crossed. We provide an analytical proof and numerical evidence that the anomaly is invariant under independent SO(3) rotations of $\mathbf{E}$ and $\mathbf{B}$.

We numerically solve the lattice Weyl fermion model in the presence of a strong magnetic field and find that the longitudinal magnetoconductivity is a nonuniversal quantity, depending on the direction of the magnetic field, microscopic parameters, and nonlinear dispersion. Using an analytical solution developed for a special case, we find that nonlinear dispersion induces a vacuum shift that modifies the magnetic-field dependence of the magnetoconductivity. These results indicate that magnetoconductivity alone is not a reliable signal of the chiral anomaly. Instead, we find that the magnetoconductivities factorize into a product of an anomaly-related universal part and a density-of-states (DOS)-related part.

Based on this observation, we propose a rotationally invariant observable,
\ba
\varkappa = \sigma \, c_V^2 / T^2,
\ea
where $\sigma = \sqrt{\sigma_L^2 + \sigma_H^2}$ is the Euclidean norm of the longitudinal and Hall conductivities, and $c_V = C_{N,V}/V$ is the specific heat density. One can show that at low temperatures, $c_V/T \propto \varrho/V,$ with $\varrho(\mu)$ the DOS at the Fermi level. This quantity is remarkably insensitive to dispersion details and exhibits a universal $B^2$ dependence. Importantly, we demonstrate that $\varkappa \propto |\cos\Theta|$, where $\Theta$ is the angle between $\mathbf{E}$ and $\mathbf{B}$, with the longitudinal and Hall components satisfying $\sigma_L \varrho^2 \propto B^2 \cos^2\Theta$ and $\sigma_H \varrho^2 \propto B^2 \sin\Theta \cos\Theta$, respectively. All non-universal aspects of the conductivity are absorbed into $\varrho$, rendering $\varkappa$ a robust and unambiguous probe of the chiral anomaly in solid-state systems.

\color{blue}\emph{Model}\color{black}. --- We begin with a lattice model for Weyl fermions described by the Hamiltonian
\ba
\hat{H} = \sum_{\mathbf{k}} \hat{\psi}^{\dagger}(\mathbf{k}) \, \mathbf{d}(\mathbf{k}) \cdot \boldsymbol{\sigma} \, \hat{\psi}(\mathbf{k}),
\ea
where $\mathbf{d}(\mathbf{k}) = J_{\perp} \big( \sin k_x,\; \sin k_y,\; 2 - \cos k_x - \cos k_y + (J_z/J_\perp)(m - \cos k_z) \big)^{\mathsf{T}}, \hat{\psi}(\mathbf{k}) = (\hat{c}_{\uparrow}(\mathbf{k})$, $\hat{c}_{\downarrow}(\mathbf{k}))^{\mathsf{T}}$, and $\hat{c}_{\sigma}(\mathbf{k})$ annihilates an electron with spin $\sigma = \uparrow$, $\downarrow$ at momentum $\mathbf{k}$. The matrices $\boldsymbol{\sigma} = (\sigma_x, \sigma_y, \sigma_z)$ are the Pauli matrices. For $|m| < 1$ and $J_\perp > J_z$, the bands touch at two Weyl points $\mathbf{K}_{\pm} = (0, 0, \pm K^*)$ with $K^* = \arccos m$. Expanding around $\mathbf{K}_{\pm}$ to linear order in $\mathbf{p} = \mathbf{k} - \mathbf{K}_{\pm}$, we obtain the low-energy Hamiltonian
\ba
H(\mathbf{p}) = J_\perp p_x \Gamma_x + J_\perp p_y \Gamma_y + J_z \sin K^* \, p_z \Gamma_z,
\ea 
where $\Gamma_{x,y} = \sigma_{x,y} \otimes I$ and $\Gamma_z = \sigma_z \otimes \tau_z$ form a Clifford algebra, with $\tau_i$ Pauli matrices acting in valley space. In accordance with the Nielsen–Ninomiya theorem \cite{NielsenNinomiya1981}, Weyl fermions appear in pairs of opposite chirality, so that the chiral anomaly manifests as inter-valley charge transfer. Crucially, Lorentz and SO(3) rotational symmetries are absent at the microscopic level. Even at low energies, this symmetry is explicitly broken because generically $J_\perp \neq J_z \sin K^*$. In what follows, we demonstrate the robustness of the anomaly form, the effect of nonlinear dispersion on longitudinal magnetoresistance, and the universality of $\varkappa$ in the presence of short-range disorder.

\color{blue}\emph{Robustness of the Chiral Anomaly}\color{black}.--- We consider a magnetic field $\mathbf{B}$ with direction $\hat{\mathbf{B}} = (\sin\theta\sin\phi,\; \sin\theta\cos\phi,\; \cos\theta)$ (so that $\mathbf{B}$ lies in the $y$-$z$ plane when $\phi=0$). Minimal coupling is implemented via $\mathbf{k} \rightarrow \mathbf{k} - \frac{e}{\hbar}\mathbf{A}$, where the vector potential is chosen in a gauge analogous to the Landau gauge:
\ba
\mathbf{A} = B\big( \sin\phi\,x'\cos\theta,\; \cos\phi\,x'\cos\theta,\; -x'\sin\theta \big),
\ea
where $x' = x\cos\phi - y\sin\phi$, $y' = x\sin\phi + y\cos\phi$. This transformation can be written as $\mathbf{r}' = \hat{R}_{\hat{z}}(\phi)\mathbf{r}$, where $\hat{R}_{\hat{n}}(\phi)=e^{i\phi\hat{n}\cdot\hat{\mathbf{I}}}$ denotes a rotation by angle $\phi$ about axis $\hat{n}$. Correspondingly, the unit vectors transform as $\hat{\mathbf{e}}' = \hat{R}_{\hat{z}}(\phi)\hat{\mathbf{e}}$, and the momentum operator transforms as $\hat{\mathbf{k}} = \hat{R}_{\hat{z}}(-\phi)\hat{\mathbf{k}}'$. In the rotated frame, the vector potential becomes $\mathbf{A}' = B(0,\; x'\cos\theta,\; -x'\sin\theta)$, and we set $\mathbf{k}' \rightarrow \mathbf{k}' - \frac{e}{\hbar}\mathbf{A}' \equiv \vec{K}'$.

In this frame, the $\mathbf{d}$-vector takes the form
\ba
\mathbf{d}'(\mathbf{k}',\hat{x}') &=& J_\perp\left( \sin(\vec{K}'\cdot\hat{e}'_x), \sin(\vec{K}'\cdot\hat{e}'_y), 2\! -\! \cos(\vec{K}'\cdot\hat{e}'_x) \right.\nonumber\\
&&\left.- \frac{mJ_z}{J_\perp} - \cos(\vec{K}'\cdot\hat{e}'_y) - \frac{J_z}{J_\perp}\cos(\vec{K}'\cdot\hat{e}_z) \right),
\ea
where, for example, $\vec{K}'\cdot\hat{e}'_x = \cos\phi\,\hat{k}'_x + \sin\phi\,k'_y - e\cos\theta\sin\phi\,\hat{x}'$ (the other components are obtained analogously). In this gauge, $k'_y$ and $k'_z$ are conserved quantities, while $\hat{k}'_x = -i\partial_{x'}$ and $\hat{x}'$ are conjugate variables.

We now perform an additional rotation about the $\hat{e}'_x$-axis, defined by $\mathbf{k}'' = \hat{R}_{\hat{e}'_x}(\theta)\mathbf{k}'$. In this new coordinate system, $\hat{e}''_z$ aligns with $\hat{\mathbf{B}}$, while $\hat{e}'_x = \hat{e}''_x$ remains unchanged. The vector potential simplifies to $\mathbf{A}'' = B(0, x', 0)$, and the $\mathbf{d}$-vector becomes
\ba
\mathbf{d}'(\vec{K}'') &=& J_\perp\left( \sin(\vec{K}''\cdot\hat{e}'_x),\sin(\vec{K}''\cdot\hat{e}''_y),2 \!-\! \cos(\vec{K}''\cdot\hat{e}'_x)\right.\nonumber\\
&&\left. - \frac{mJ_z}{J_\perp} - \cos(\vec{K}''\cdot\hat{e}''_y) - \frac{J_z}{J_\perp}\cos(\vec{K}''\cdot\hat{e}''_z) \right),
\ea
with $\vec{K}'' = (\hat{k}'_x,\; k''_y - \hat{x}'/\ell_B^2,\; k''_z)^\mathsf{T}$ and $\ell_B^2 = \hbar/eB$.
\begin{figure}[t]
\hspace{-0ex}\includegraphics[width=8.4cm]{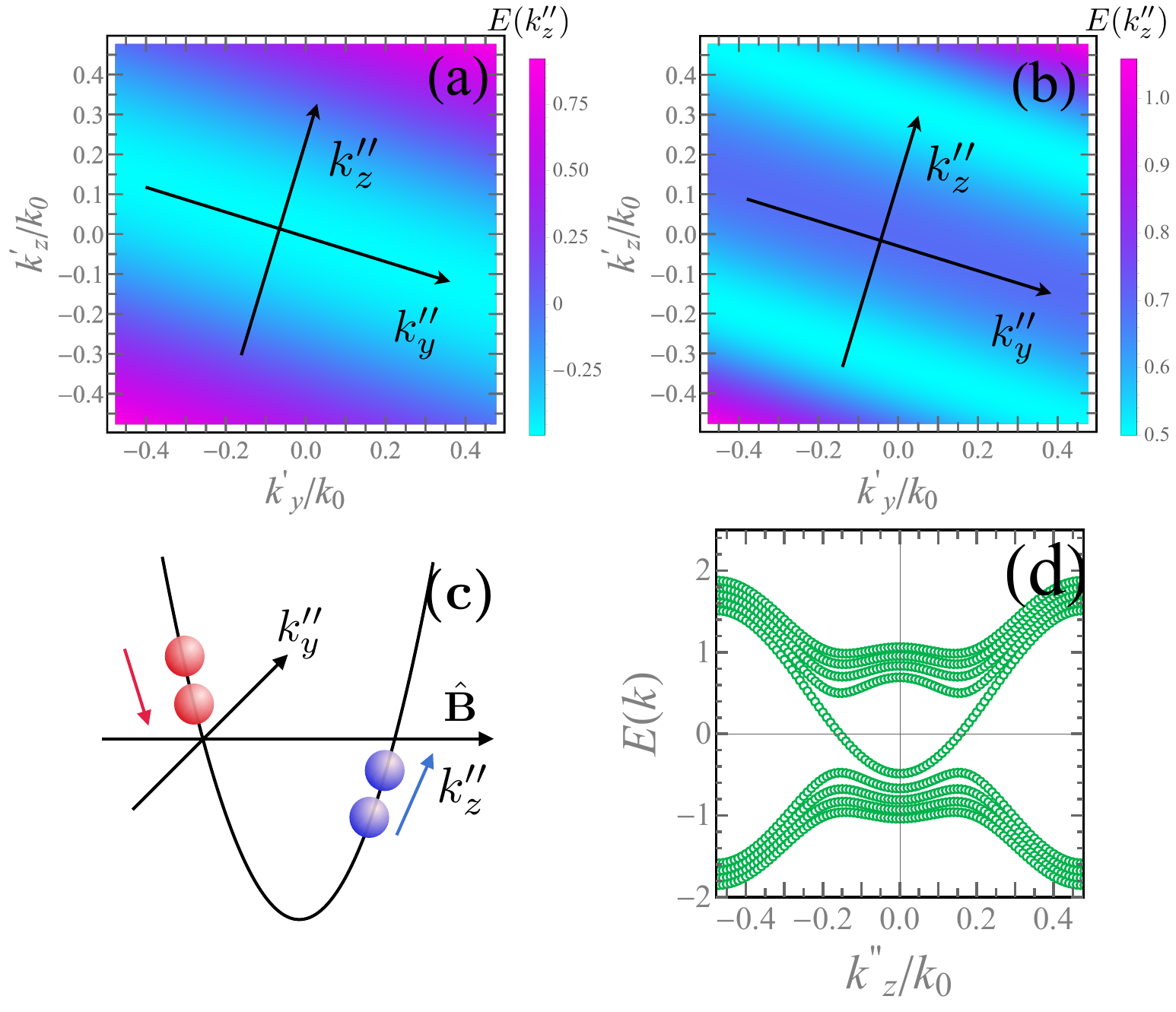}
\caption{a,b:The eigenvalue $E_n(k_y',k_z)$ of different band. ($n=0$(a) is the chiral band, $n=1$(b) is the first band above chiral band). We fix $J_z=1,J_{\perp}=5,m=0.5,\Theta=0.1\pi,\Phi=0.25\pi,\nu_B=0.005$, and fix the cut of operator $\hat{a}$ to be 400 in the occupation representation. c:Schematic illustration of the chiral anomaly mechanism in the lattice. d:Band dispersion along the magnetic field direction.   }
\label{FigLL}
\end{figure}

This construction yields three key consequences:
\begin{enumerate}
\item The quantization condition $k''_y = 2\pi n_{y''}/L_{y''}$ (with $n_y \in \mathbb{Z}^+$) gives the Landau level degeneracy $\mathcal{D} = (L_{x'}/\ell_B^2)/(2\pi/L_{y''}) = eB L_{y''} L_{x'}/2\pi\hbar$.
\item Defining $\hat{x}'' = \hat{x}' - \ell_B^2 k''_y$ reduces the problem to a one-dimensional Landau system parameterized by $k''_z$, with energies $E_n(k''_z)$ where n is the Landau level index. The lowest Landau level $E_0(k''_z)$ is chiral in the sense that it has only one branch in the vicinity of the Weyl nodes.
\item An electric field $\mathbf{E}$ can be introduced via $\tilde{\mathbf{A}} = \mathbf{E}t$. This additional vector potential affects only the parameter $k''_z$, since changes in $k''_y$ affect the degeneracy and changes in $k''_x$ correspond to band projection. Noting that $k''_z$ is directed along $\hat{\mathbf{B}}$, the electric field shifts $E_0(k''_z)$ to $E_0\big(k''_z - (e/\hbar)(\mathbf{E}\cdot\hat{\mathbf{B}})\,t\big)$.
\end{enumerate}

We now consider the regime where the Fermi level crosses only the chiral Landau band. This occurs for moderate magnetic fields: weaker fields lead to multiple Landau levels crossing the Fermi energy (semiclassical regime), while extremely strong fields can induce Weyl node annihilation \cite{Zhang2017}. In this quantum limit, the chiral anomaly rate is given by
\ba
\frac{d(N_R - N_L)}{dt} = \mathcal{D} \rho_0 \frac{dE_0}{dt} = \mathcal{D} \rho_0 \frac{\partial E_0}{\partial k''_z} \frac{d}{dt}\!\left(k''_z - \frac{e}{\hbar}\mathbf{E}\cdot\hat{\mathbf{B}}\,t\right),
\ea
where $\rho_0$ is the one-dimensional density of states for the band $E_0(k''_z)$, and $v_F = \partial E_0/\partial k''_z$ is the Fermi velocity. For any one-dimensional system, $\rho_0 v_F = L''_z/\pi$ (with a factor of 2 accounting for the two Weyl nodes). The Landau level degeneracy is $\mathcal{D} = S_{x''y''} eB/2\pi\hbar$. Substituting these expressions yields
\ba
\partial_\mu J^\mu_5 = \frac{1}{V}\frac{d(N_L - N_R)}{dt} = \frac{e^2 B}{2\pi^2\hbar^2}\,\mathbf{E}\cdot\hat{\mathbf{B}} = \frac{e^2}{2\pi^2\hbar^2}\,\mathbf{E}\cdot\mathbf{B}.
\ea
Thus, the chiral anomaly retains its universal form even in lattice Weyl fermions with strongly modified and anisotropic dispersion. The only assumption required is that the lattice constants are equal in all three directions. This demonstrates that the emergent SO(3) symmetry of the anomaly originates from ultraviolet physics and reflects the topological nature of charge conservation.

\color{blue}{\emph{Weyl Fermions in Strong Magnetic Fields and Nonlinear Dispersion}}\color{black}. --- Using the gauge introduced above, we numerically solve the Landau level problem starting from Eq.~\ref{Eq:LL}. Periodic boundary conditions are imposed along the $\hat{e}'_x$ direction. Since $\hat{k}'_x$ and $\hat{x}'$ are conjugate variables, we introduce ladder operators via $\hat{k}'_x = -i(\hat{a} - \hat{a}^\dagger)/(\sqrt{2}\ell_B)$ and $\hat{x}' = \ell_B(\hat{a} + \hat{a}^\dagger)/\sqrt{2}$. The momentum operator $\vec{K}'(\hat{k}'_x, k'_y - \ell_B^2\cos\theta\,\hat{x}', k'_z + \ell_B^2\sin\theta\,\hat{x}')$ then becomes an expression in $\hat{a}$ and $\hat{a}^\dagger$, and the Hamiltonian transforms accordingly. The eigenenergies are obtained numerically; the two-dimensional energy spectra for different bands are shown in the insets of Fig.~\ref{FigLL}(a) and (b). For all bands, we observe a degeneracy along the direction defined by $\hat{k}''_y = \cos\theta\sin\phi\,\hat{k}_x + \cos\theta\cos\phi\,\hat{k}_y - \sin\theta\,\hat{k}_z$ \cite{Supplementary}. The direction orthogonal to $k''_y$ corresponds to $k''_z$, i.e., the direction of $\hat{\mathbf{B}}$. The resulting band dispersions $E_n(k''_z)$ are presented in Fig.~\ref{FigLL}(d).

To illustrate the solution in the moderate-field regime, we consider the simplified case $\theta = 0$. Introducing ladder operators satisfying $[\hat{x}, \hat{k}_x] = i$ with $\hat{x} = \ell_B(\hat{a} + \hat{a}^\dagger)/\sqrt{2}$ and $\hat{k}_x = -i\ell_B^{-1}(\hat{a} - \hat{a}^\dagger)/\sqrt{2}$, and applying the Baker–Hausdorff expansion, we obtain
\ba
\cos(\hat{k}_x) = \frac{1}{2} e^{-\nu_B/4} \left( e^{-\hat{a}^\dagger / \sqrt{2/\nu_B}} e^{\hat{a} / \sqrt{2/\nu_B}} + \mathrm{h.c.} \right),
\ea
where $\nu_B = \lambda^2 / \ell_B^2$ and the factor $e^{-\nu_B/4}$ arises from the vacuum shift due to nonlinear dispersion. In the quantum limit $\nu_B \ll 1$ (where only one Landau level crosses the Fermi energy), the effective Hamiltonian reduces to
\ba
\hat{H}(\mathbf{k}) = i\tilde{J}_\perp^B (\hat{a}\sigma^+ - \hat{a}^\dagger\sigma^-) + \left( \tilde{E}_0(k_z) + \frac{\sqrt{\nu_B}\tilde{J}_\perp^B}{\sqrt{2}} \hat{a}^\dagger\hat{a} \right) \sigma_z,
\ea
with $\tilde{J}_\perp^B = \sqrt{\nu_B}J_\perp e^{-\nu_B/4}/\sqrt{2}$ and $\tilde{E}_0(k_z) = 2J_\perp(1 - e^{-\nu_B/4}) + J_z(m - \cos k_z)$. The chiral mode $|\Omega\rangle = |n_a=0, \uparrow, k_z\rangle$ (satisfying $\hat{a}|\Omega\rangle = \sigma^-|\Omega\rangle = 0$) yields the density of states
\ba
\frac{\varrho(\mu)}{V} = \frac{eB}{2\pi} \left| J_z^2 - J_\perp^2 \left( 2(1 - e^{-\nu_B/4}) + \frac{mJ_z - \mu}{J_\perp} \right)^2 \right|^{-1/2}. \label{DOS}
\ea
Notably, in addition to the linear $B$ dependence arising from the Landau level degeneracy—which has been identified as a source of nonmonotonic Hall response in Weyl semimetals \cite{Nagaosa24}—the density of states also acquires a $B$ dependence through $\nu_B$, reflecting the vacuum shift induced by nonlinear dispersion. As we will show below, this additional B dependence contributes to the non-universal part of the magnetoconductivity.

\color{blue}\emph{Longitudinal and Hall Magnetoconductivity}\color{black}. --- For short-range disorder characterized by $\overline{V(\mathbf{r})V(\mathbf{r}')} = \gamma^2 \delta(\mathbf{r} - \mathbf{r}')$, we employ the Kubo formalism. The conductivity tensor is given by
\ba
\sigma_{\alpha\beta} = \lim_{\omega \to 0} \frac{i}{\omega + i0^+} \left[ K_{\alpha\beta}(\omega) - K_{\alpha\beta}(0) \right],
\ea
with
\ba
K_{\alpha\beta}(\omega) = \int d\epsilon \, d\epsilon' \, \frac{n_F(\epsilon') - n_F(\epsilon)}{\epsilon' - \epsilon + \omega + i0^+} \, \mathrm{Tr} \left[ \hat{J}_\alpha A(\epsilon) \hat{J}_\beta A(\epsilon') \right],
\ea
where $\hat{J}_\alpha = e \, \partial \mathcal{H}(\vec{K}')/\partial k_\alpha$ is the current operator along direction $\alpha$, and $A(\epsilon) = \frac{1}{\pi} \mathrm{Im}(\epsilon - \hat{H} - i\tau_0^{-1})^{-1}$ is the spectral function matrix.

Within the self-consistent Born approximation (SCBA) and neglecting anti-weak-localization corrections, the scattering rate takes the form
\ba
\tau_0^{-1} = \pi \gamma^2 \, (\varrho(\mu)/V), \label{BornA}
\ea
where $\varrho(\mu)$ is the total density of states at the Fermi energy $\mu$. Since $\varrho(\mu)$ itself depends on $\tau_0$, Eq.~\ref{BornA} must be solved self-consistently.

Using these expressions, we perform numerical calculations via exact diagonalization of $\hat{H}$. For the longitudinal conductivity $\sigma_L$, we set $\hat{e}_\alpha = \hat{e}_\beta = \hat{\mathbf{E}}$; for the Hall conductivity $\sigma_H$, we take $\hat{e}_\alpha = \hat{\mathbf{E}}$ and $\hat{e}_\beta = \hat{\perp} = \hat{\mathbf{E}} \times (\hat{\mathbf{E}} \times \hat{\mathbf{B}})$. Numerical results are presented in Figs.~2(a)–2(d). In Fig.~2(a), we compare $\sigma_L(\nu_B)$ obtained from the linearized dispersion with that from the full lattice model, revealing a vacuum shift arising from the density-of-states modification discussed in Eq.~\ref{DOS}. Figure~2(b) shows the dependence of $\sigma_L$ on the magnetic field direction angles $\theta$ and $\phi$, while Figs.~2(c) and 2(d) display $\sigma_L$ and $\sigma_H$ as functions of $\Theta = \angle(\hat{\mathbf{E}}, \hat{\mathbf{B}})$ for various microscopic parameters $J_\perp/J_z$ and electric field orientations. The raw data exhibit non-universal behavior; the calculations are performed for $\nu_B = 0.020$ and $0.025$, both lying within the quantum limit.
\begin{figure}[t]
\includegraphics[width=8.6cm]{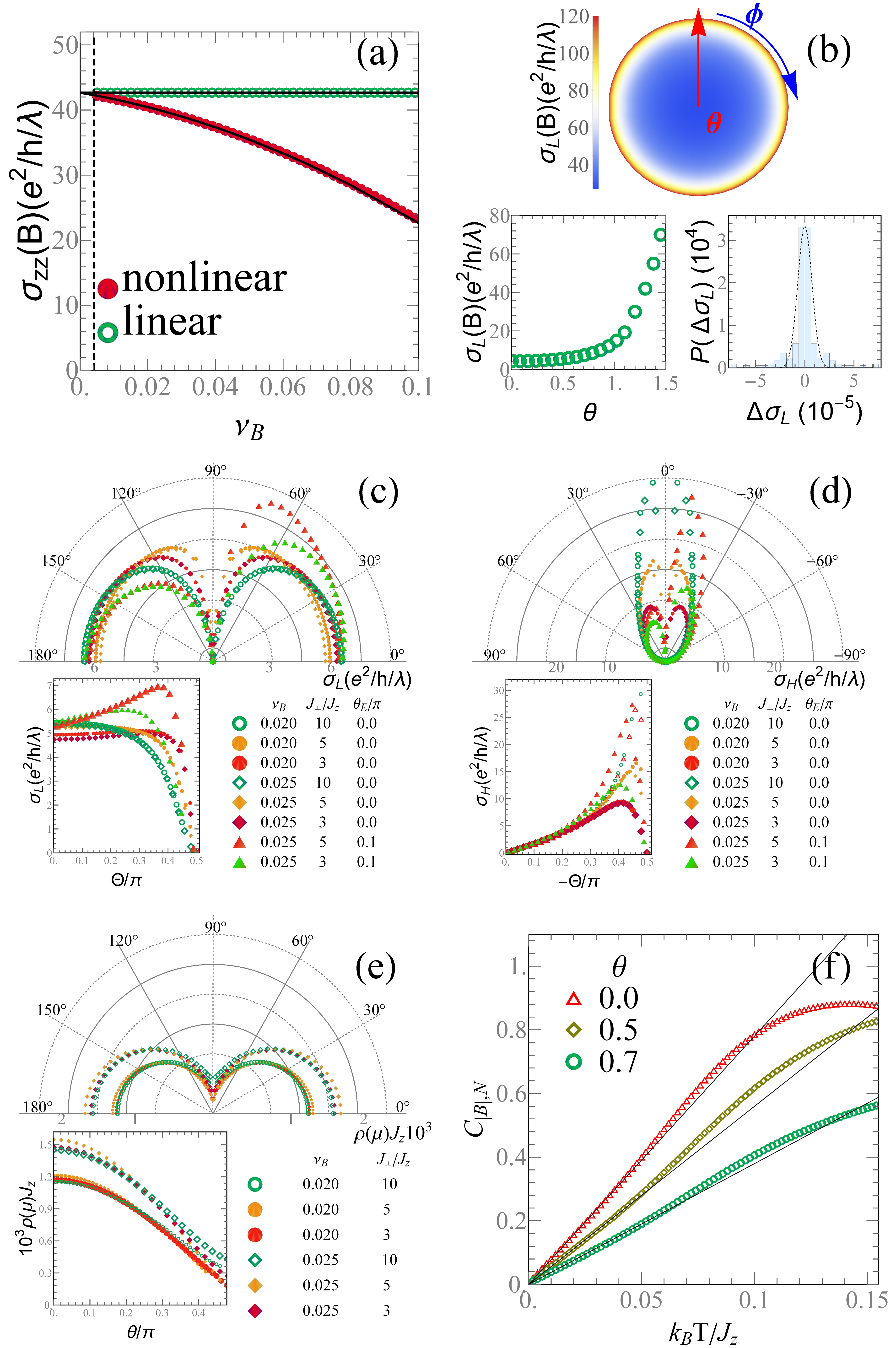}
\caption{(a):We fix $\mu/J_z=0.3,m=0.5,\gamma/J_z=0.3,J_x/J_z=10$.The relationship between longitudinal conductivity and magnetic field strength when both electric and magnetic fields are aligned in the z-direction. The real line is obtained through analytical approximation. The qualitative difference between the blue and black lines reflects the correction to conductivity induced by vacuum shift due to nonlinear effects. (b): The relationship between longitudinal conductivity and the direction of the magnetic field $\hat{\mathbf{B}}=(\sin{\theta}\cos{\phi},\sin{\theta}\sin{\phi},\cos{\theta})$, while maintaining $\angle(\hat{\bf B},\hat{\bf E})=\pi/6$. (c,d): The $\sigma_L$ and $\sigma_H$ as functions of $\Theta$ under different parameters. We  keep the direction of the magnetic field in the y-z plane . (e):The Dos as a function of $\theta$. (f): The temperature dependence of low-temperature heat capacity $C_{N,V}(T)$.  We fix $\nu_B=0.02$ in Fig.b,f and $\mu/J_z=0.05,m=0.5,\gamma/J_z=0.3$ in (b) - (f). }
\label{Num}
\end{figure}
\begin{figure}[t]
\includegraphics[width=8.6cm]{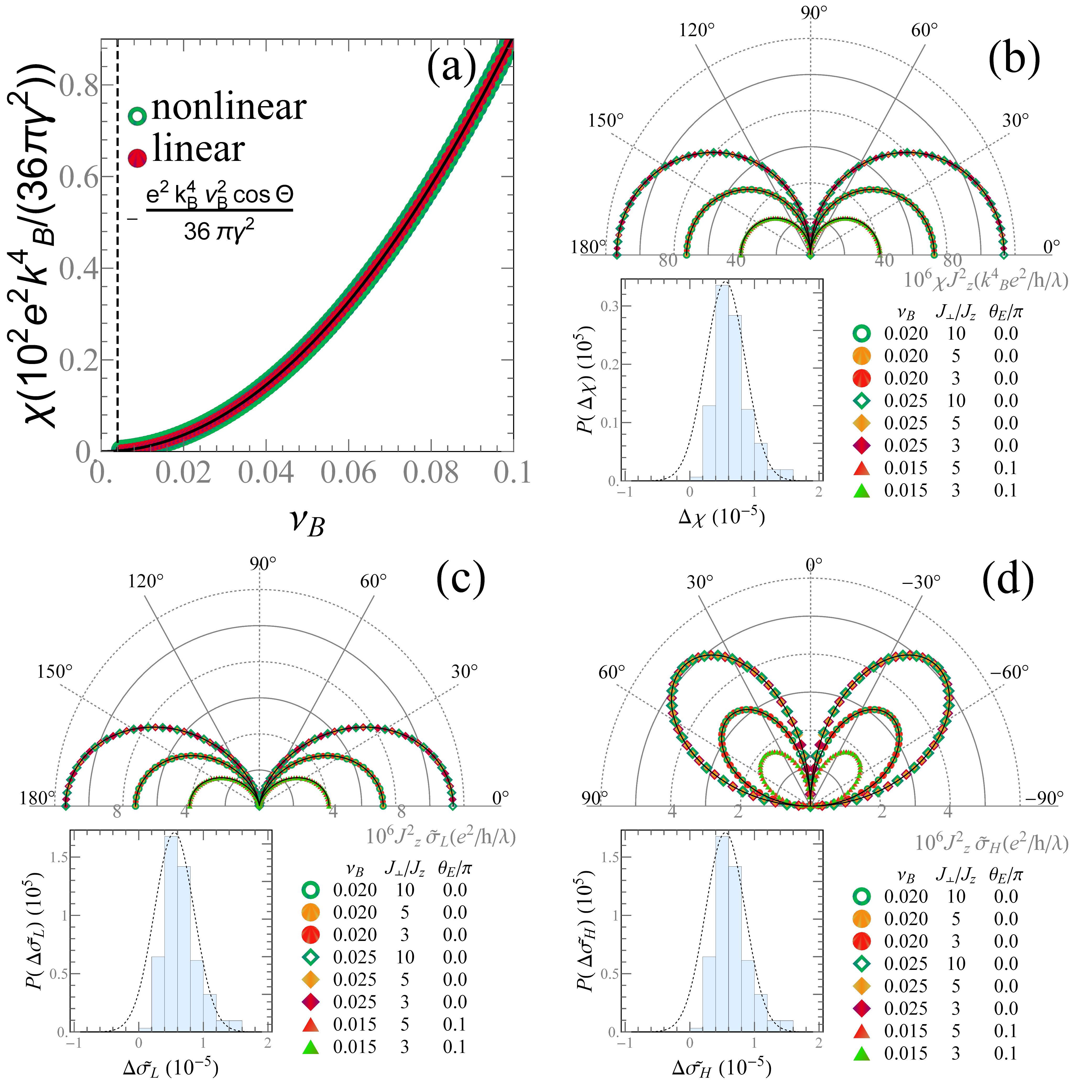}
\caption{(a):The figure demonstrates that the same result for $\varkappa$ is obtained regardless of whether a linear approximation is made near the Weyl points or not. (b,c,d): We fix $\mu/J_z=0.05,m=0.5,\gamma/J_z=0.3$, and Keep the direction of the electric field in the z direction and the magnetic field in the y-z plane . These figure shows that $\varkappa,\tilde{\sigma}_H,\tilde{\sigma}_L$ falls into the same function of $\Theta$ for different $J_{\perp}/J_z$. The inset displays a statistical histogram of the relative errors between numerical calculations and analytical approximate expressions.}
\label{Top}
\end{figure}

We now turn to an analytical calculation of $\sigma_L$ and $\sigma_H$. We begin by factorizing the Hamiltonian as $\hat{H} = \sum_n E_n(k''_z) \hat{P}_{n,k''_y,k''_z}$, where $\hat{P}_{n,k''_y,k''_z} = |\psi_{n,k''_y,k''_z}\rangle\langle \psi_{n,k''_y,k''_z}|$ projects onto the $n$-th Landau level with momenta $k''_y, k''_z$. Here $|\psi_{n,k''_y,k''_z}\rangle$ denotes the eigenstate of $\hat{H}$ in a magnetic field. Inserting this decomposition into the current operator yields
\ba
J_{\hat{\mathbf{E}}} = (\hat{\mathbf{E}} \cdot \hat{e}''_z) \sum_{n,k''_y} \left( \frac{\partial E_n(k''_z)}{\partial k''_z} \hat{P} + E_n(k''_z) \frac{\partial \hat{P}}{\partial k''_z} \right).
\ea
Using the identities $\hat{P}^2 = \hat{P}$, $\mathrm{Tr}(\hat{P} \partial_{k''_z}\hat{P}) = \partial_{k''_z} \mathrm{Tr}(\hat{P}^2) = 0$, and $\mathrm{Tr}(\partial_{k''_z}\hat{P} \, \hat{P} \partial_{k''_z}\hat{P} \hat{P}) = 0$, we obtain
\ba
\sigma_L &=& \frac{e^2}{V} \sum_n \int d\epsilon \, \delta(\epsilon - \mu) \, (\hat{e}''_z \cdot \hat{\mathbf{E}})^2 \, \mathrm{Tr} \left[ \left( \frac{\partial E_n(k''_z)}{\partial k''_z} \right)^2\right.\nonumber\\
&&\left. \hat{P}_{n,k''_y,k''_z} \left( \frac{\tau_0^{-1}}{(\epsilon - E_n(k''_z))^2 + \tau_0^{-2}} \right)^2 \right],
\ea
where the trace includes summation over $k''_y$, $k''_z$, and spin degrees of freedom.

In the quantum limit, the sum over n is restricted to the chiral Landau level only. Noting that $\hat{e}''_z = \hat{\mathbf{B}}$, we find
\ba
\sigma_L &\approx& \frac{e^2}{V} \int d\epsilon \, \delta(\epsilon - \mu) \, (\hat{\mathbf{B}} \cdot \hat{\mathbf{E}})^2 \, \mathrm{Tr} \left[ \left( \frac{\partial E_0(k''_z)}{\partial k''_z} \right)^2\right.\nonumber\\
&&\left. \hat{P}_{0,k''_y,k''_z} \left( \frac{\tau_0^{-1}}{(\epsilon - E_0(k''_z))^2 + \tau_0^{-2}} \right)^2 \right].
\ea
Evaluating the trace gives
\ba
\sigma_L = e^2 (\hat{\mathbf{B}} \cdot \hat{\mathbf{E}})^2 \, \mathcal{D} \sum_{k''_z} (v''_z)^2 \, \delta(\mu - E_0(k''_z)) \, \tau_0,
\ea
with $\mathcal{D} = eB S/\hbar$ the Landau level degeneracy. This simplifies to
\ba
\sigma_L \approx e^2 (\hat{\mathbf{E}} \cdot \hat{\mathbf{B}})^2 \, (v_F^2 \tau_0) \, (\varrho(\mu)/V).
\ea
A similar calculation for the Hall conductivity yields
\ba
\sigma_H \approx e^2 (\hat{\mathbf{E}} \cdot \hat{\mathbf{B}})(\hat{\mathbf{B}} \cdot \hat{\perp}) \, (v_F^2 \tau_0) \, (\varrho(\mu)/V),
\ea
where $\hat{\perp}$ is the unit vector perpendicular to both $\mathbf{E}$ and $\mathbf{B} \times \mathbf{E}$. In the Born approximation (neglecting anti-weak-localization), $\tau_0^{-1} \propto \varrho(\mu)$.

The conductivities thus factorize into universal geometric factors—$(\hat{\mathbf{E}} \cdot \hat{\mathbf{B}})^2$ for $\sigma_L $ and $(\hat{\mathbf{E}} \cdot \hat{\mathbf{B}})(\hat{\mathbf{B}} \cdot \hat{\perp})$ for $\sigma_H$—multiplied by a non-universal part proportional to $v_F^2$. Using $v_F = L''_z/\pi \varrho_0 = L''_z \mathcal{D}/\pi \varrho_0 \mathcal{D} = (eB/2\pi^2\hbar)(V/\varrho(\mu))$, we see that the non-universal part depends only on the density of states. Consequently, the anomaly manifests in the universal geometric structure, while all material-specific details are absorbed into $\varrho(\mu)$.

To eliminate this DOS dependence, we note that the specific heat at constant particle number and volume, $C_{N,V}(T)$, is proportional to the DOS at the Fermi level at low temperatures. Explicitly,
\ba
c_V = \frac{C_N(T)}{V} = \frac{\pi^2}{3} \left( \frac{\varrho(\mu)}{V} \right) k_B^2 T.
\ea
Numerical verification of this relation is provided in Figs.~2(e) and 2(f), which show the DOS density $\varrho(\mu)/V$ and the specific heat density $C_{N,V}/V$, respectively.

\color{blue}{\emph{Universal Observable $\varkappa$ respecting chiral anomaly}}\color{black}. --- Given the decomposition of longitudinal (\(\sigma_L\)) and Hall (\(\sigma_H\)) conductivities in the quantum regime into universal geometric factors and DOS contributions, we find \(\sigma_{L/H}\times(\rho(\mu)/V)^2\) are universal. We find
\ba
\varkappa = \sqrt{\sigma_L^2 + \sigma_H^2}  \left(\frac{c_V}{T}\right)^2=\frac{\pi e^2  k_B^4 B}{9\times \gamma^2 E}\left|\frac{e^2{\bf E}\cdot{\bf B}}{2\pi^2\hbar^2}\right|,
\ea  
where the $B^2$ dependence and $|\hat{\bf E}\cdot\hat{\bf B}|$ dependence are the finger print of the chiral anomaly. 
As shown in Fig.~3, the rescaled conductivities \(\tilde{\sigma}_L = \sigma_L (\varrho/V)\) and \(\tilde{\sigma}_H = \sigma_H (\varrho/V)^2\) collapse onto universal curves across variations in \(\hat{\mathbf{B}}\), \(\hat{\mathbf{E}}\), \(J_z/J_\perp\), \(J_\perp\), and \(m\). The \(|\mathbf{B}|\) and \(\Theta\) (angle between \(\mathbf{E}\) and \(\mathbf{B}\)) dependencies of \(\varkappa\) further confirm universality.  (The universality is not limited by $J_x=J_y=J_\perp$, it applies for $J_x\neq J_y$ as wel l\cite{Supplementary}.)

\color{blue}\emph{Summary and Outlook}\color{black}. --- In this work, we give a proof for chiral anomaly' form being invariant in Lattice Weyl fermions with non-linear dispersion distortion in quantum region ( When magnetic field strength is large enough that the Fermi energy cut only the chiral Landau band). A SO(3) symmetry is presented when the inferred SO(3) symmetry is explicitly broken. We obtained the analytical formula for the longitudinal and Hall magneto-conductivities, and find a universal quantity combined with specific heat density $(C_{N,V}/V)$ in diffusive region. These universal quantities show SO(3) symmetry which is absent in microscopic model, manifesting the signal of chiral anomaly in its invariant form.  These analytical results are verified by numerical simulations with clear data clasps.

\emph{Acknowledgment}. --- This work is supported by the National Natural Science Foundation of China (Grants No. 12174358), the National Key R\&D Program of China (Grant No. 2022YFA1405302), and NSAF (Grant No. U2330401).

\end{document}